\documentclass[fleqn,usenatbib]{mnras}

\usepackage[T1]{fontenc}

\DeclareRobustCommand{\VAN}[3]{#2}
\let\VANthebibliography\thebibliography
\def\thebibliography{\DeclareRobustCommand{\VAN}[3]{##3}\VANthebibliography}

\usepackage{graphicx}	
\usepackage{amsmath}	
\usepackage{amssymb}	
\usepackage{txfonts}

\newcommand{\update}{}
\newcommand{\OLambdamassEbetafree}{0.54$\pm$0.28}
\newcommand{\betafree}{0.41$\pm$0.09}
\newcommand{\betaLocalRel}{0.4036$\pm$0.0016}
\newcommand{\betaWithHz}{0.4045$\pm$0.0016}
\newcommand{\DzerocovaryBeta}{(26.06\beta-5.54)\pm0.005}
\newcommand{\OLambdamassEbetacovary}{0.675$\pm$0.079}
\newcommand{\OLambdamassEbetacovaryPCIGALE}{0.678$\pm$0.078}

\newcommand{\wimproveCMB}{1.3}
\newcommand{\wimproveSN}{1.7}
\newcommand{\wimproveBAO}{2.2}

\newcommand{\lnDErrcomb}{0.34\%}
\newcommand{\lnDErrStat}{0.17\%}
\newcommand{\lnDErrBeta}{0.3\%}
\newcommand{\medianlnDA}{6.1987$\pm$0.0034}

\title[Sub-percentage distance measure at $z$=0.1]{Sub-percentage measure of distances to redshift of 0.1 by a
  new cosmic ruler}

\author[Y. Shi et al.]{
Yong Shi,$^{1,2}$\thanks{E-mail: yshipku@gmail.com}
Yanmei Chen,$^{1,2}$
Shude Mao,$^{3,4}$
Qiusheng Gu,$^{1,2}$
Tao Wang,$^{1,2}$
Xiaoyang Xia,$^{5}$
Zhi-Yu Zhang.$^{1,2}$
\\
$^{1}$School of Astronomy and Space Science, Nanjing University, Nanjing 210093, China.\\
$^{2}$Key Laboratory of Modern Astronomy and Astrophysics (Nanjing University), Ministry of Education, Nanjing 210093, China.\\
$^{3}$Department of Astronomy, Tsinghua University, Beijing 100084, China.\\
$^{4}$National Astronomical Observatories, Chinese Academy of Sciences, 20A Datun Road, Chaoyang District, Beijing 100012, China.\\
$^{5}$Tianjin Astrophysics Center, Tianjin Normal University, Tianjin 300387, People's Republic of China.\\
}

\date{Accepted XXX. Received YYY; in original form ZZZ}

\pubyear{2021}

\begin{document}
\label{firstpage}
\pagerange{\pageref{firstpage}--\pageref{lastpage}}
\maketitle

\begin{abstract}

   Distance-redshift diagrams probe expansion history of the Universe.
   We show  that the stellar  mass-binding energy (massE)  relation of
   galaxies proposed in our previous study offers a new distance ruler
   at cosmic scales.
   By using elliptical galaxies in  the main galaxy
   sample of the Sloan Digital Sky Survey Data Release 7, we construct
   a distance-redshift  diagram over the  redshift range from 0.05 to
   0.2 with the  massE ruler. The  best-fit dark energy
   density is \OLambdamassEbetacovary\, for flat $\Lambda$CDM,
   consistent with those by other  probes.  At the
   median redshift of 0.11, the median distance is estimated  to have
   a fractional error of
   \lnDErrcomb,  much  lower  than  those by  supernova  (SN)  Ia  and
   baryonic  acoustic  oscillation  (BAO) and  even  exceeding  their
   future capability at this  redshift.  The above low-$z$ measurement
   is  useful  for   probing  dark  energy  that   dominates  at the late
   Universe. For a flat dark energy equation of state  model  (flat $w$CDM), the
   massE alone  constrains $w$ to  an error that  is only a  factor of
   \wimproveBAO, \wimproveSN\, and \wimproveCMB\ times  larger than those by
   BAO, SN Ia, and cosmic microwave background (CMB), respectively.

   
\end{abstract}




\section{Introduction}

A  distance-redshift  diagram  is  a  powerful  probe  of  the  cosmic
expansion history. The measurement by E.  Hubble and G.  Lema{\^\i}tre
revealed the  expansion of the Universe  \citep{Hubble29, Lemaitre27},
laying  down the  foundation  of the  big bang  theory.   In the  past
decades,  two  independent  distance  rulers  at  cosmic  scales  with
appreciable  redshifts, supernova  Ia  (SN Ia)  and baryonic  acoustic
oscillation (BAO), demonstrated unambiguously the acceleration of the
cosmic   expansion,   implying   the    existence   of  dark   energy
\citep{Riess98, Perlmutter99, Eisenstein05}.

It  is  essential  to  have independent  rulers  to  measure the distance
vs. redshift  diagram. Physically, different rulers  probe the cosmic
geometry  from   different  aspects,   thus  examining   a  concordant
cosmological   model.   Technically,   each  ruler   has  ``nuisance''
parameters that are  unrelated to the cosmological  parameters but can
introduce  systematic errors  \citep[e.g.][]{Zuntz15}.  Joint
analysis by combining different rulers can significantly alleviate the
degeneracy among cosmological parameters and improve their accuracy.

Uncertainties of current distance measurements using SN Ia and BAO are
around a  few percent and vary  with redshift, mainly limited  by the
sample size  and survey volume \citep{Ross15,  Alam17, Scolnic18}.  In
the forthcoming  5-10 years, large  imaging and spectroscopic  surveys will
reach precision from a few sub-percent to one percent per redshift bin
of  0.1 \citep{Feng14,  DESI16}.  In  this letter,  we  show that  the
relation between  stellar masses and  binding energies of  galaxies as
proposed in \citet{Shi21} offers distance rulers at cosmic scales with
significant statistical power. Through the study we will refer  this relation as massE where ``mass''
stands for stellar masses and ``E'' for binding energies.

\section{The massE cosmic ruler}

\citet{Shi21} demonstrated  a tight relation between  stellar mass and
binding energy of  galaxies, which covers nine orders  of magnitude in
stellar masses  over a large range  of galaxy types and  redshift. The
overall  relationship  is described  by  double  power laws  with  the
transition galaxy stellar mass  of $\sim$ 10$^{7}$-10$^{8}$ M$_{\odot}$.
As illustrated in their Fig. 2  and  3, the dependence on redshift
is absent up to $z$ $\sim$ 2.5, so does  the dependence on galaxy properties
such  as  galaxy  sizes,  surface  densities and star  formation  rates.
 The physical
origin of the massE relation may  be related to the self-regulation of
a  galaxy between  its binding  energy and  the accumulative  feedback
energy  from its  stellar  populations.  Although  the massE  relation
holds  for various  types  of  galaxies, in  this  study  we only  use
elliptical galaxies  so that the  velocity dispersion can  be measured
through single fibers.

For  galaxies   with  stellar  masses  well   above  10$^{8}$
M$_{\odot}$, the massE relation can be written as a single power law:
\begin{equation}\label{eqn_relation}
  \sigma_{\rm  e}R_{\rm e}^{0.25}   = AM_{\rm  star}^{\beta},
\end{equation}
 where $R_{\rm e}$ is the effective radius  that encloses half of stellar light, $M_{\rm star}$ is the total
stellar masses of galaxies, $\sigma_{\rm  e}$ is the velocity dispersion within $R_{\rm e}$ for
dispersion-dominated galaxies, and $\beta$=\betaLocalRel\, (see \S~\ref{appendix_massElocal}). $\sigma_{\rm e}R_{\rm e}^{0.25}$
represents the fourth root of the galaxy binding energies within $R_{\rm e}$.

In  Equation~\ref{eqn_relation}, $\sigma_{\rm  e}$ is independent  of
redshift,   $\frac{R_{\rm   e}}{\rm kpc}$=$\frac{R_{\rm   e,as}}{206.27}$$\frac{D_{\rm A}}{\rm Mpc}$  and
$M_{\rm star}$=$M_{\rm star,1Mpc}$$(\frac{D_{\rm  L}}{\rm 1 Mpc})^{2}$, where $R_{\rm  e,as}$ is
the apparent  effective radius in  arcsec, $M_{\rm star,1Mpc}$  is the galaxy
stellar mass placed at  a distance of 1  Mpc, the angular
diameter  distance  $D_{\rm A}$  is  related to  the  luminosity  distance
$D_{\rm L}$ in $D_{\rm A}$=$D_{\rm L}$/(1+$z$)$^{2}$. $M_{\rm star,1Mpc}$ can be
obtained by first redshifting the observed spectral energy distribution
(SED) to the rest-frame and then fit with stellar population synthesis
models to obtain the mass-to-light ratio and the corresponding mass at
a distance of 1 Mpc. $M_{\rm star,1Mpc}$ is essentially a flux-like quantity and its measurement
is independent of cosmological models.

As a result, the corresponding angular diameter distance calculator is:
\begin{eqnarray}\label{distance_ruler}
  \frac{D_{\rm A}}{\rm Mpc} & = &  S_{D_{0}} \frac{D_{0}}{\rm Mpc}\left[(1+z)^{-4\beta}\frac{\sigma_{\rm e}}{\rm km\,s^{-1}}(\frac{M_{\rm star,1Mpc}}{M_{\odot}})^{-\beta}\right.  \nonumber   \\ 
  & &  \left.(\frac{R_{\rm e,as}}{\rm arcsec})^{0.25}\right]^{1/(2\beta-0.25)},  
\end{eqnarray}
with  $D_{0}$=($\frac{0.264~{\rm  km~s^{-1}~kpc^{0.25}~M_{\odot}^{-1}}
}{A})^{1/(2\beta-0.25)}$ in  Mpc.   When   applying
Equation~\ref{distance_ruler}   to   estimate    distances   for 
cosmological purpose, the exact $D_{0}$  always degenerates with the
local Hubble  constant $H_{0}$,  similar to the  case of  the absolute
magnitude of SN  Ia. As a result, a nuisance  parameter $S_{D_{0}}$ is
included as  the scaling factor  for a  sample to reproduce  the local
Hubble constant, and  is related to the absolute  calibration of three
observables ($\sigma_{\rm  e}$, $M_{\rm star,1Mpc}$,  $R_{\rm e,as}$).
The  second nuisance  parameter $\beta$  can be  estimated from  local
samples as detailed in \S~\ref{appendix_massElocal}.

Not  that  although   the  massE  relation  invokes   the  same  three
observables  as  the  fundamental   plane,  they  are  different  both
physically and quantitatively. As shown in \citet{Shi21}, the massE is
a correlation between galaxy stellar masses and binding energies while
the fundamental  plane is  between galaxy  stellar mass  and dynamical
masses.  For  galaxies above 10$^{8}$ $M_{\odot}$,  the massE relation
only  has   two  free   parameters  that  are   $A$  and   $\beta$  in
Equation~\ref{eqn_relation},  while the  fundamental  plane has  three
free  parameters  that  are  $a$,  $b$   and  $c$  in  Equation  5  of
\citet{Shi21}.  Furthermore, it  has been  known that  the fundamental
plane needs  a fourth hidden  systematic parameter to account  for the
dependence    on    galaxy    properties   or    local    environments
\citep[e.g.][]{Bernardi03, Magoulas12,  Howlett22}.  As a  result, the
fundamental plane  is mainly  limited to z  $<$ 0.05-0.1  for peculiar
velocity studies \citep[e.g.][]{Howlett22}.

The   following  strategy   is  proposed   to  measure   the  distance
vs. redshift diagram using the massE cosmic ruler.  For a selected sample
in a given redshift bin, its median redshift, if ignoring the peculiar
velocity,    corresponds exactly to its  median  luminosity  distance,
because  the  luminosity   distance  increases
monotonically with redshift, independent  of cosmological models.  As a
result, we  measure the  median of the  redshift distribution  and the
median of the luminosity  distance distribution in individual redshift
bins  to  obtain  the  distance  vs.   redshift  diagram.   With  this
strategy,  the sample  incompleteness does  not matter,  i.e., objects
rejected  by  the  selection  function  does not  affect  at  all  the
correspondence between  the median  distance and median  redshift. But
the Malmquist bias  does matter in a  way that, e.g., a  sample with a
magnitude cut is  intrinsically fainter so that their  stellar mass is
over-estimated.

\begin{figure}
  \centering
  \includegraphics[scale=0.28]{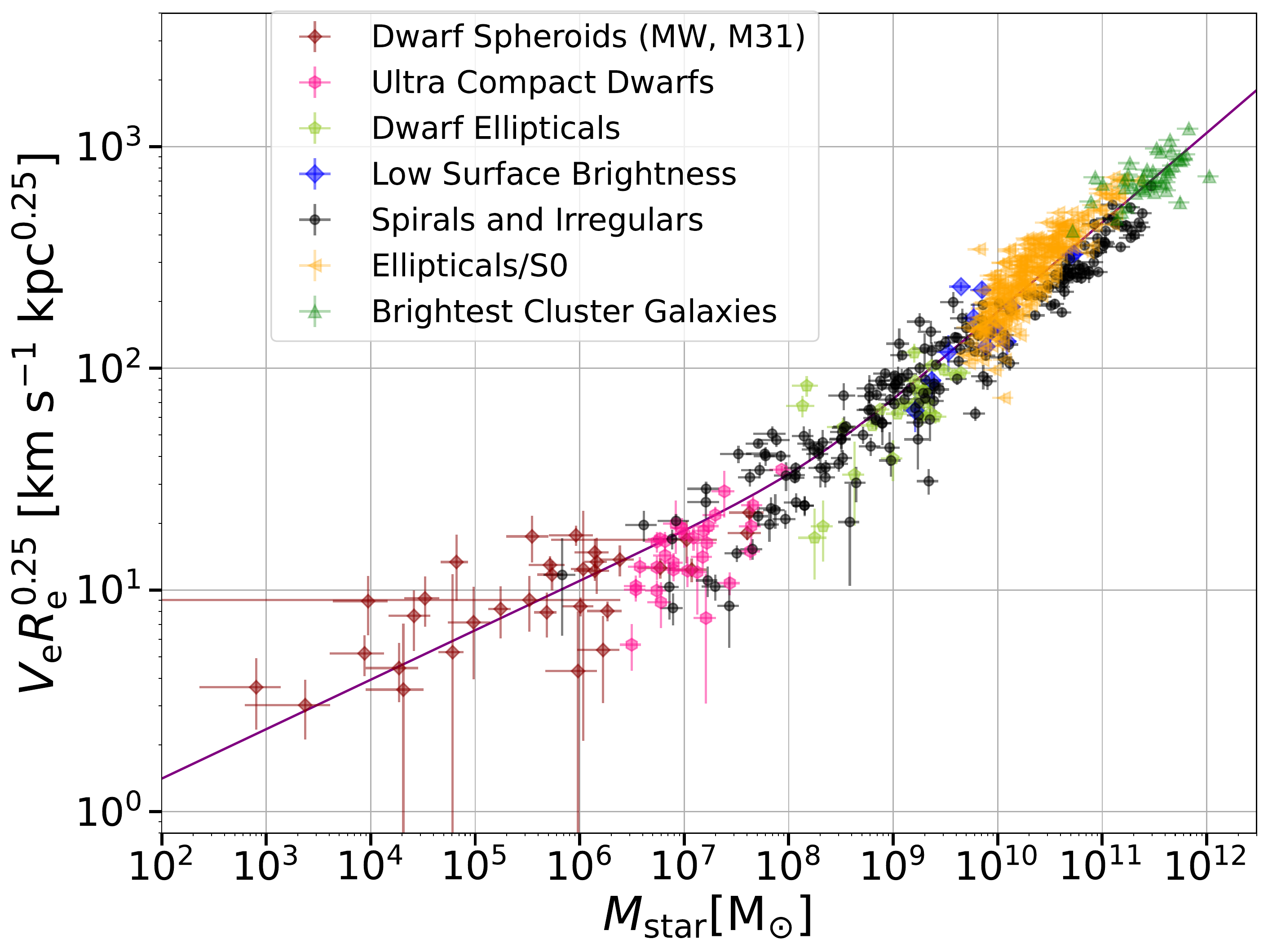}
  \caption{\label{local_massE} The local sample compiled in \citet{Shi21} that is used to
    derive the intrinsic massE relationship. $V_{\rm e}$ is rotation velocity and velocity
    dispersion for rotation dominated and dispersion dominated galaxies, respectively. For details, see \citet{Shi21}. The solid line indicates the derived intrinsic relationship after considering
    errors on both X and Y axes (see \S~\ref{appendix_massElocal}). }
\end{figure}

\begin{figure*}
  \centering
  \includegraphics[scale=0.53]{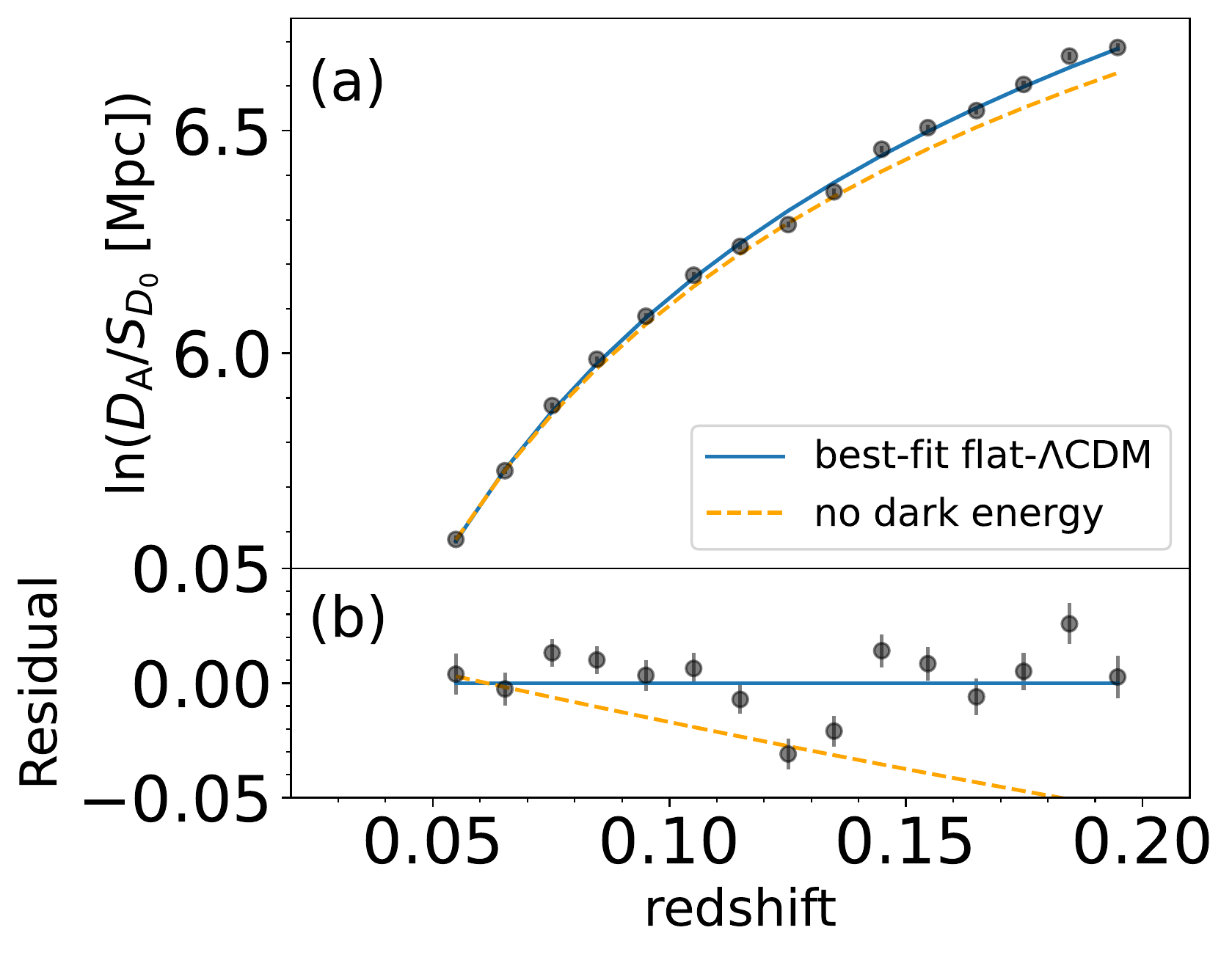}
   \raisebox{-0.3cm}{\includegraphics[scale=0.53]{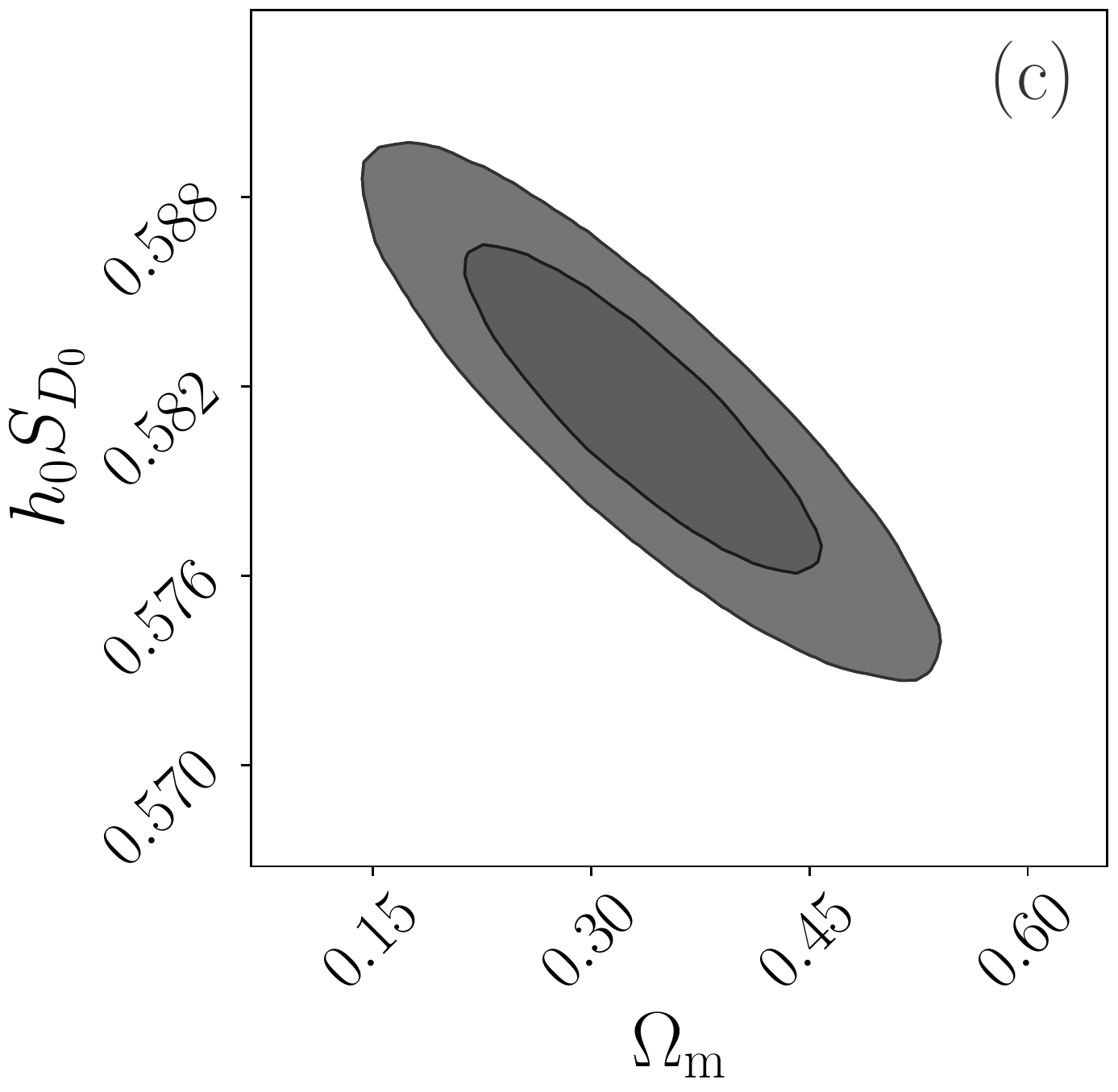}}
   \caption{\label{lnDA_vs_z_SDSS} {\bf (a)} The distance vs. redshift diagram that is derived from elliptical galaxies
     in the SDSS main galaxy sample using the massE ruler. $D_{\rm A}$
     is the angular-diameter distance in Mpc and $S_{D_{0}}$ is the scale factor
     that calibrates the absolute distance and degenerates with local Hubble
     constant. The error bars indicate the square root of the diagonal elements
     of the covariance matrix shown in in Figure~\ref{covariance_matrix} (a). 
     The solid line represents the best-fit flat $\Lambda$CDM with $\Omega_{\Lambda}$=\OLambdamassEbetacovary.
     The dashed line is the one without dark energy normalized at two lowest-redshift bins. {\bf (b)} The residual
     of the best fit. As shown in \S~\ref{sec_dist_error}, the scatter in the residual is consistent
     with the  covariance matrix in Figure~\ref{covariance_matrix} (a). {\bf (c)}  The confidence ranges of dark matter density ($\Omega_{\rm m}$) and
     the product $h_{0}S_{D_{0}}$ at 68\% and 95\%, where $h_{0}$ is the local Hubble constant
     in 100 km/s/Mpc. The plot was produced through
     \texttt{ChainConsumer} \citep{Hinton16}. }
\end{figure*}

\begin{figure}
  \centering
  \includegraphics[scale=0.38]{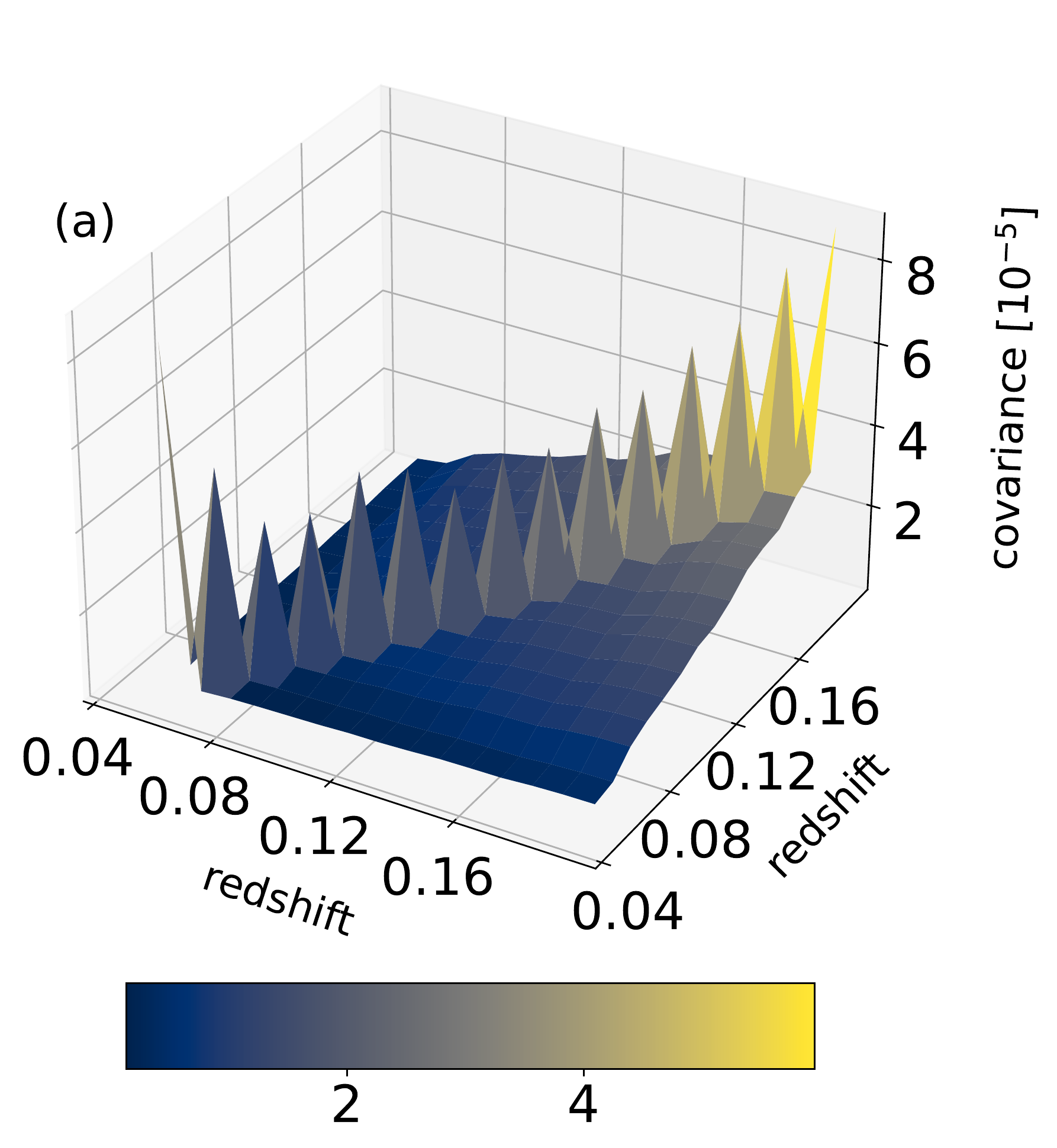}
    \includegraphics[scale=0.41]{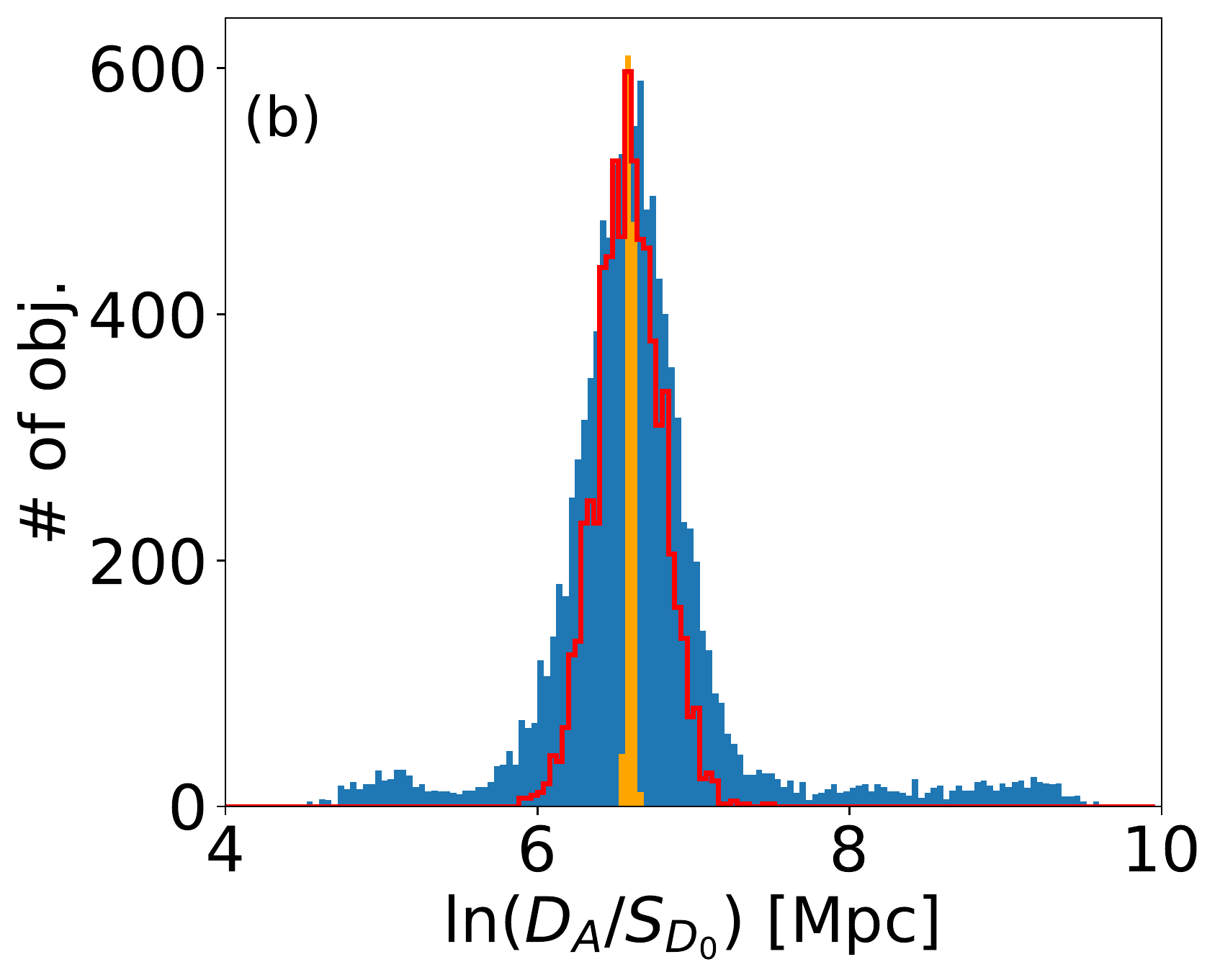}
   \caption{\label{covariance_matrix} (a) The covariance matrix of the distance measurements at different redshift bins. (b)  The distribution of
   the derived angular diameter distances of elliptical galaxies in the SDSS main galaxy sample (blue filled) within the redshift bin
    of [0.17, 0.18]. 
    The red open histogram shows the simulated distributions based on the median errors of three observables in the galaxy catalog, and the orange filled histogram
    is the simulation that only considers the observed redshift distribution plus random peculiar  velocities of 1000 km/s (orange filled). Note that the two simulated
  distributions are rescaled to have  the same median and histogram peak as the observed one.  }
\end{figure}

\section{The massE relationship of the local sample}\label{appendix_massElocal}

Before  applying the  massE  to  the SDSS  sample,  we  first need  to
estimate  the  intrinsic  massE relationship,  especially  $\beta$  in
Equation~\ref{eqn_relation}.   We  thus  perform the  fitting  to  the
sample  compiled and  homogenized  in  \citet{Shi21}, while  excluding
luminous infrared galaxies and ultra diffuse galaxies because of their
large observational  errors as  well as high-$z$  sources in  order to
remove  the  dependence  of   the  relationship  on  the  cosmological
model. This local sample contains 589 objects as shown in Fig.~\ref{local_massE}. Measurements of galaxy  sizes, velocity dispersions and stellar
masses of  the sample in  general have high signal  to noise. We
thus  add  additional  0.1  dex,  10\%  and  1\%  systematic  errors
quadratically  to $M_{\rm  star}$, $R_{\rm  e}$ and  $\sigma_{\rm e}$,
respectively.   The  systematic error  for  $M_{\rm  star}$ is  mainly
caused by uncertainties in population synthetic modeling, which can be
illustrated by comparing the MPA-JHU and CIGALE stellar masses of SDSS
galaxies in \S~\ref{sec_flatCDM}.  The  uncertainty for $R_{\rm e}$ is
from  background subtraction,  morphology irregularity,  galaxy center
determination etc.  We estimated this by selecting a subsample of SDSS
galaxies  with high  $R_{\rm  e}$ signal-to-noise  ratio  $>$ 100  and
comparing measurements  from different  data releases.  The systematic
error for  $\sigma_{\rm e}$  is mainly  caused by  absolute wavelength
calibration and 1\% is a reasonable estimate for modern CCD-based spectra \citep{Law21}. If increasing the errors by a
small  amount, e.g.,  to  0.15 dex  for $M_{\rm  star}$,  to 15\%  for
$R_{\rm e}$ and  to 2\% for $\sigma_{\rm e}$,  respectively, the difference
in the derived slope $\beta$ is within one $\sigma$.

We  performed  PyMC3
\citep{Salvatier16} by  considering errors on both  $M_{\rm star}$ and
$\sigma_{\rm e}R_{\rm e}^{0.25}$ to obtain the intrinsic relationship.
The best fit gives $\beta$=\betaLocalRel\, and the corresponding $D_{0}$  co-varies with
$\beta$ through:
\begin{equation}\label{eqn_D0_beta}
  {\rm ln} ({D_{0}}/{\rm Mpc})={ \DzerocovaryBeta}.
\end{equation}

 The derived   slope  $\beta$   is  similar   to  that   in
\citet{Shi21} but  with a smaller  error, because in  \citet{Shi21} the
fitting was done without  weighting by observational errors, which thus gives the result including the
observational error instead of the intrinsic relationship. Distances
of this local sample are mostly  below 40 Mpc, with the remaining 17\%
of the  sample distributes between  40 and  300 Mpc.  They  are mainly
based on cosmological-independent measurements  for those below $\sim$
20 Mpc,  but are  homogenized to a  flat $\Lambda$CDM  with $H_{0}$=73
km/s/Mpc,  and   $\Omega_{\Lambda}$=0.73  for  those  above   20  Mpc.
However, the best-fit  $\beta$ changes only by a  small fraction of
its  1-$\sigma$   error  if  changing  to   $\Omega_{\Lambda}$=1.0  or
$\Omega_{\Lambda}$=0.4, respectively.  This  is because the associated
change in  $R_{\rm e}$ and $M_{\rm  star}$ is much smaller  than their
observational  errors, plus  the  fact  that both  X  and Y-axes  vary
simultaneously to partially cancel out the effect.

We also carry out additional fitting to the massE relationship by including
105 high-$z$ objects compiled in \citet{Shi21}. The derived
$\beta$=\betaWithHz\, is essentially the same as the one with only the local sample.
This suggests that $\beta$ should not evolve strongly with the redshift.

\section{Elliptical  galaxies in the  main galaxy sample of  Sloan Digital
  Sky Survey (SDSS)}

Although all types of galaxies can be used for
the massE  ruler, elliptical galaxies  are the most suitable  ones given
their low extinction, regular morphologies, low current star formation
rates  and dispersion dominated  kinematics, which  will facilitate
measurements of  three observables  ($\sigma_{\rm e}$,  $M_{\rm star,1Mpc}$,
$R_{\rm  e,as}$).  Unlike  spiral   galaxies  whose  velocity  requires
spatially-resolved   kinematic   maps,   ellipticals'  can   rely   on
single-fiber  measurements \citep{Cappellari06,  Shi21}.  We  build  our
elliptical    galaxy    sample    from   the    value-added    MPA-JHU
catalog \citep{Kauffmann03,  Brinchmann04}  of  the  SDSS  data  release
7 \footnote{https://wwwmpa.mpa-garching.mpg.de/SDSS/DR7/}.
A sample of  about 28,0000 elliptical galaxies were  selected from the
SDSS main galaxy  sample with the following criteria:  (1) no redshift
warning  as indicated  by \texttt{Z\_WARNING}=0;  (2) the  main galaxy
sample as  defined by \texttt{PRIMTARGET}  $\geq$ 64 but $<$  128 and
Petrosion magnitude  in $r$ $\leq$  17.77; (3) elliptical  galaxies as
defined by \texttt{FRACDEV} $\geq$ 0.8. Only the second criterion will introduce
the Malmquist bias while others are not related to the three observables ($\sigma_{\rm e}$,
$R_{\rm e,as}$, $M_{\rm star,1Mpc}$).

To  apply  Equation~\ref{distance_ruler},  we adopt the  Petrosian
radius in the $r$ band (\texttt{PETROR50\_R})  to represent $R_{\rm e}$.  The total stellar
mass $M_{\rm  star}$ adopts  \texttt{MSTAR\_TOT} that is  derived from
fitting  to Petrosian  magnitudes  in four  SDSS  optical bands.   The
advantage of  the Petrosian radius and  magnitude is that there is no  bias in
redshift due  to the surface  dimming \citep{Blanton01}, which  is the
key for the distance vs. redshift measurement.  $M_{\rm  star}$ is converted to
$M_{\rm  star,1Mpc}$ simply through the adopted cosmology for the MPA-JHU catalog.

We carry out our own  measurements of the velocity dispersion, because
 those in  the MPA-JHU catalog were  derived by using
the   median   instrumental  resolution,   which  will   introduce
redshift-dependent    bias as the true instrumental  resolution is a function of the wavelength.      The    spectrum     and    associated
wavelength-dependent  instrumental  resolution  of  each  object  were
available                 in                  the                 SDSS
archive\footnote{http://data.sdss.org/sas/dr12/sdss/spectro/redux/26/spectra/}. The
fitting was  done using the  \texttt{pPXF} code  \citep{Cappellari17} by
adopting  the  \texttt{HR-PYPOPSTAR} theoretical  stellar  populations
with 50,000 resolving  power at 5000\AA$\;$ \citep{Millan-Irigoyen21}.
During the  fitting, we  further exclude 5800-6100\AA$\;$  range where
blue and red channels of  the spectrograph with different instrumental
resolutions overlap.  The derived  velocity dispersion within  a 3$''$
fiber  was  then corrected  to  that  at \texttt{PETROR50\_R}  by  the
following equation \citep{Cappellari06}:
\begin{equation}
 \sigma_{\rm    e}     =    \sigma_{\rm     fiber}(R_{\rm    e}/R_{\rm fiber})^{0.066\pm0.035}.
\end{equation}
Here the error on the power-law index is implemented through \texttt{numpy.random.normal}.

\begin{figure*}
  \begin{center}
  \includegraphics[scale=0.6]{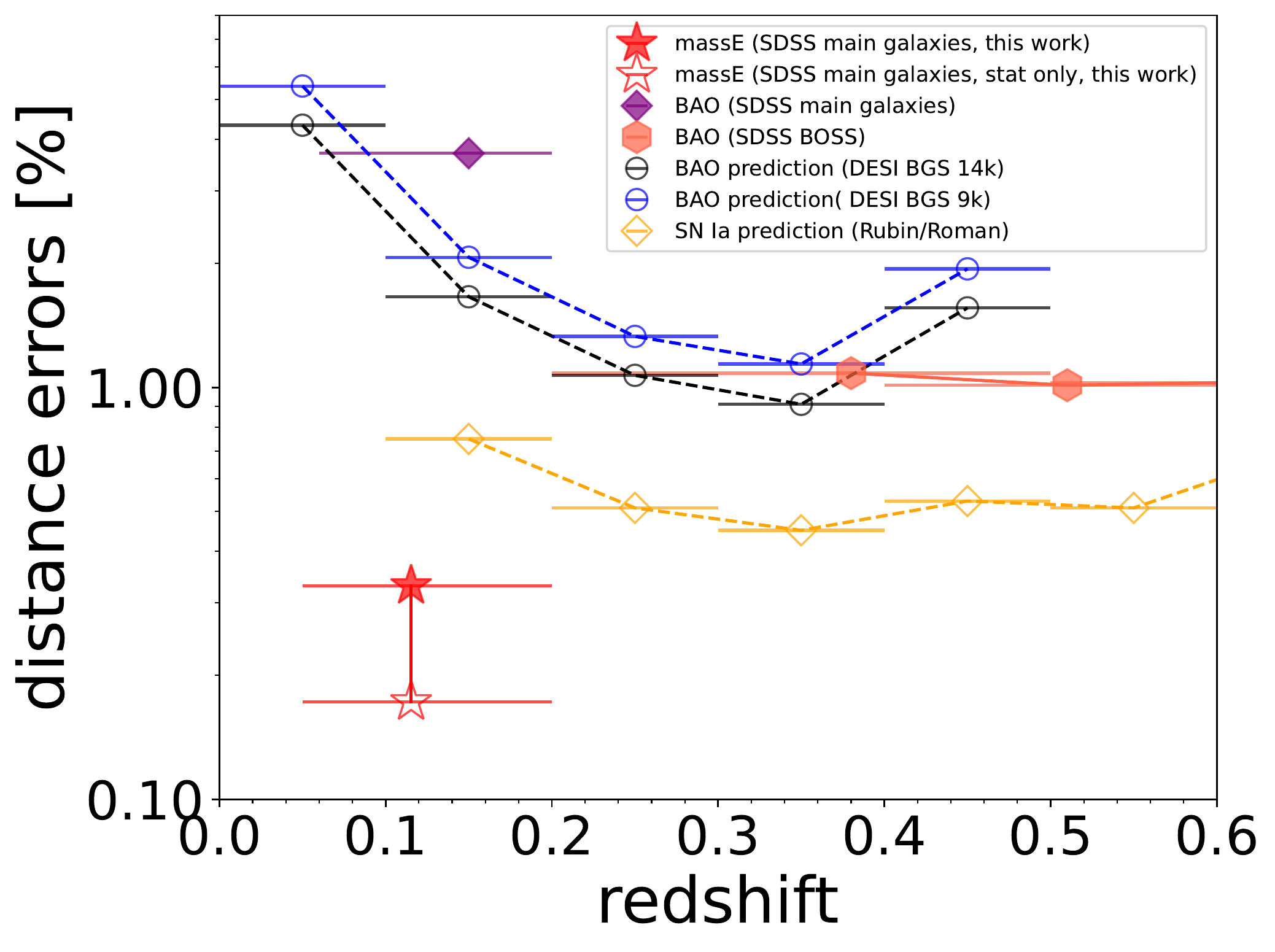}
  \caption{\label{comp_distance_error} The fractional errors of distance measurements at different redshifts by different methods, including massE with SDSS main
    galaxies (this work),
    BAO with SDSS main galaxy survey (MGS)  \citep{Ross15}, BAO with SDSS BOSS data \citep{Alam17},  BAO prediction with DESI \citep{DESI16} and  SN Ia prediction
    with Rubbin/Roman \citep{Feng14}.  }
  \end{center}
\end{figure*}

\begin{figure}
  \begin{center}
  \includegraphics[scale=0.50]{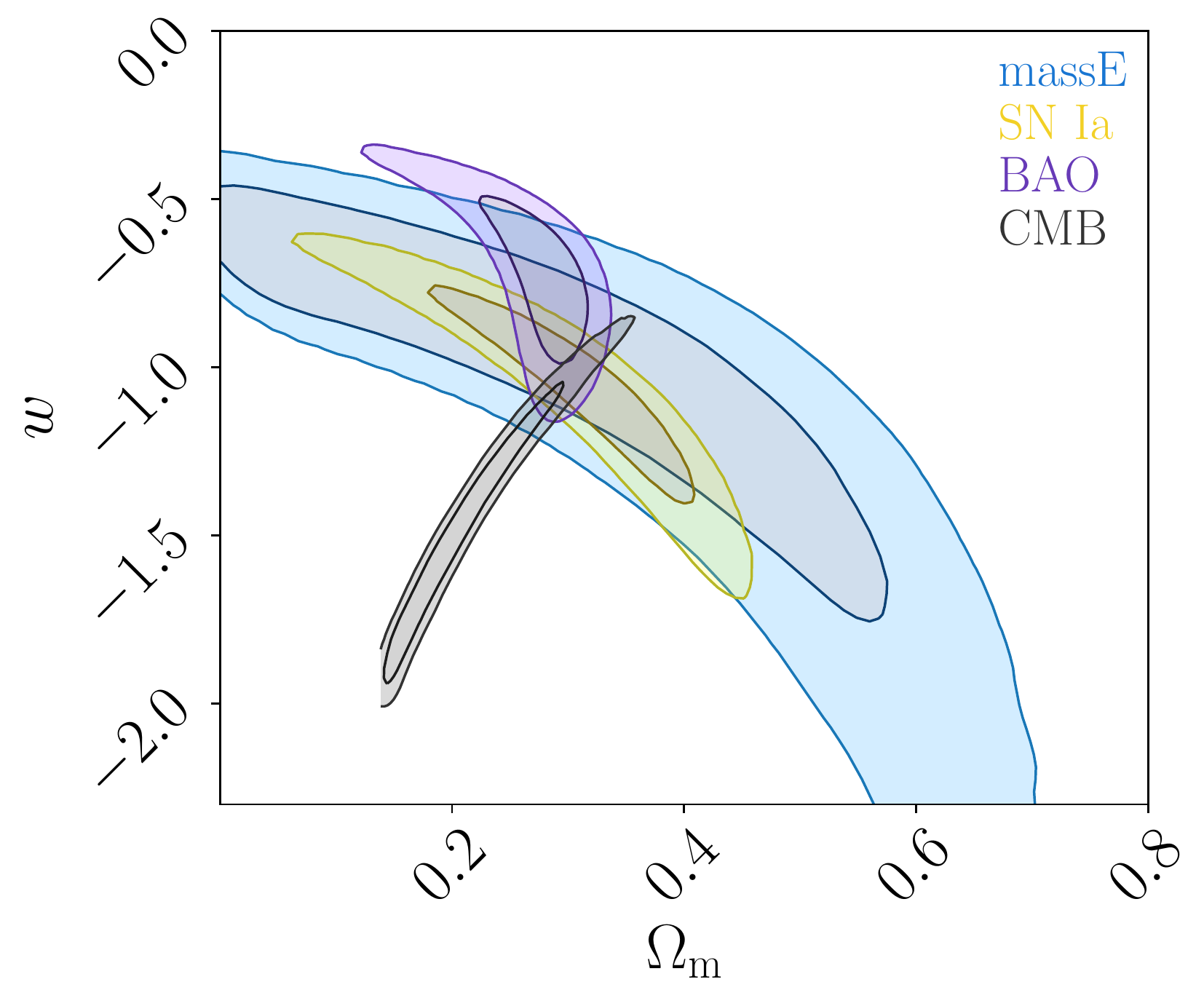}
  \caption{\label{w_omegam_DiffProbers} The confidence range of the dark energy equation of state ($w$) and the dark matter density
    ($\Omega_{\rm m}$) for the flat $w$CDM model  by CMB, BAO, SN Ia and massE, respectively. The lower-left cut for the CMB is caused
    by the $H_{0}$ prior that is set to be $<$ 100 km/s/Mpc. The plot was produced through
     \texttt{ChainConsumer} \citep{Hinton16}. }
  \end{center}
\end{figure}

\section{The  distance  vs.   redshift diagram  of  the  SDSS  elliptical
  sample.}\label{sec_dist_sdss}

\begin{table}
\caption{\label{tab_dist_result} The angular diameter distance as derived by the massE ruler.}
\begin{tabular}{llllllllllllll}
\hline
ID & $z_{\rm med}$ & ln($D_{A}(z_{\rm med})/S_{D_{0}}$) \\
\hline
            &   &  ln(Mpc) \\
\hline
1 &     0.0549 &     5.5840 \\
2 &     0.0652 &     5.7371 \\
3 &     0.0752 &     5.8830 \\
4 &     0.0847 &     5.9884 \\
5 &     0.0950 &     6.0851 \\
6 &     0.1051 &     6.1746 \\
7 &     0.1150 &     6.2391 \\
8 &     0.1251 &     6.2892 \\
9 &     0.1348 &     6.3628 \\
10 &     0.1449 &     6.4589 \\
11 &     0.1546 &     6.5070 \\
12 &     0.1649 &     6.5449 \\
13 &     0.1749 &     6.6054 \\
14 &     0.1846 &     6.6676 \\
15 &     0.1948 &     6.6847 \\

\hline
\end{tabular}\\
\end{table}

\begin{table*}
\caption{\label{tab_cov_matrix} The covariance matrix of the angular diameter distance ln($D_{A}(z_{\rm med})/S_{D_{0}}$).}
\begin{tabular}{llllllllllllllll}
\hline
 82.38 &   0.49 &   0.83 &   0.96 &   1.24 &   1.45 &   1.64 &   1.79 &   2.01 &   2.34 &   2.49 &   2.69 &   3.01 &   3.16 &   3.27 \\ 
  0.49 &  51.29 &   1.28 &   1.47 &   1.88 &   2.23 &   2.48 &   2.71 &   3.01 &   3.52 &   3.72 &   3.96 &   4.47 &   4.69 &   4.87 \\ 
  0.83 &   1.28 &  36.32 &   2.48 &   3.20 &   3.76 &   4.22 &   4.62 &   5.18 &   6.05 &   6.42 &   6.92 &   7.77 &   8.12 &   8.44 \\ 
  0.96 &   1.47 &   2.48 &  36.24 &   3.75 &   4.40 &   4.93 &   5.41 &   6.05 &   7.08 &   7.52 &   8.13 &   9.10 &   9.49 &   9.88 \\ 
  1.24 &   1.88 &   3.20 &   3.75 &  45.49 &   5.69 &   6.38 &   7.00 &   7.84 &   9.16 &   9.73 &  10.53 &  11.80 &  12.30 &  12.80 \\ 
  1.45 &   2.23 &   3.76 &   4.40 &   5.69 &  45.02 &   7.49 &   8.21 &   9.20 &  10.76 &  11.41 &  12.34 &  13.85 &  14.44 &  15.02 \\ 
  1.64 &   2.48 &   4.22 &   4.93 &   6.38 &   7.49 &  37.99 &   9.21 &  10.32 &  12.06 &  12.80 &  13.85 &  15.53 &  16.20 &  16.85 \\ 
  1.79 &   2.71 &   4.62 &   5.41 &   7.00 &   8.21 &   9.21 &  45.24 &  11.34 &  13.25 &  14.07 &  15.22 &  17.07 &  17.80 &  18.51 \\ 
  2.01 &   3.01 &   5.18 &   6.05 &   7.84 &   9.20 &  10.32 &  11.34 &  45.28 &  14.86 &  15.79 &  17.10 &  19.17 &  19.99 &  20.79 \\ 
  2.34 &   3.52 &   6.05 &   7.08 &   9.16 &  10.76 &  12.06 &  13.25 &  14.86 &  53.22 &  18.44 &  19.98 &  22.40 &  23.36 &  24.29 \\ 
  2.49 &   3.72 &   6.42 &   7.52 &   9.73 &  11.41 &  12.80 &  14.07 &  15.79 &  18.44 &  55.99 &  21.24 &  23.79 &  24.81 &  25.81 \\ 
  2.69 &   3.96 &   6.92 &   8.13 &  10.53 &  12.34 &  13.85 &  15.22 &  17.10 &  19.98 &  21.24 &  67.13 &  25.81 &  26.89 &  27.98 \\ 
  3.01 &   4.47 &   7.77 &   9.10 &  11.80 &  13.85 &  15.53 &  17.07 &  19.17 &  22.40 &  23.79 &  25.81 &  70.35 &  30.15 &  31.36 \\ 
  3.16 &   4.69 &   8.12 &   9.49 &  12.30 &  14.44 &  16.20 &  17.80 &  19.99 &  23.36 &  24.81 &  26.89 &  30.15 &  79.61 &  32.69 \\ 
  3.27 &   4.87 &   8.44 &   9.88 &  12.80 &  15.02 &  16.85 &  18.51 &  20.79 &  24.29 &  25.81 &  27.98 &  31.36 &  32.69 &  88.80 \\ 
\hline
\end{tabular}\\
All numbers are in 10$^{-6}$ ln(Mpc)$^{2}$. Each number gives the covariance
between $i$-th and $j$-th redshift bins as listed in
Table~\ref{tab_dist_result}.
\end{table*}

The redshift distribution  of the SDSS elliptical  sample peaks around
0.11.  We  limit the redshift range  to [0.05, 0.2] and  construct the
distance vs.  redshift diagram with a  redshift bin size of 0.01.  The
number of  galaxies in individual  redshift bins result  in negligible
fractional errors of their median redshift,  which is $<$ 0.06\% for a
random peculiar velocity of 1000 km/s.   The upper range at $z$=0.2 is
also necessary to avoid the Malmquist bias caused by the magnitude cut
of  $r$ $<$  17.77. Although  the massE  ruler is  not subject  to the
sample completeness,  objects that  are selected  into the  sample can
systematically  bias   the  stellar   mass  if  their   luminosity  is
overestimated due to  the Malmquist bias.  To estimate  it roughly, we
use the $r$-band luminosity  function \citep{Blanton03} to construct a
mock $r$-band  photometric sample  down to  the SDSS  5-$\sigma$ depth
($r$=22.7)  at different  redshifts, and  add the  photometric error
based  on the  median  and standard  deviation  of apparent  magnitude
errors as a function of magnitude  from the SDSS sample. By assuming a
constant mass-to-light  ratio in  the $r$-band, it  is found  that the
median  distance is  under-estimated by  1\% at  ($z$,$\Delta$$z$)=(0.2,
0.01) while by as much as  7\% at ($z$,$\Delta$$z$)=(0.25, 0.01) for the
sub-sample with $r$ $<$ 17.77. As a  result, we limit the $z$ range to
be below  0.2 so that  the Malmquist bias is  a small fraction  of the
typical 1-$\sigma$  uncertainty in  individual redshift bins  that are
around 1.3\% including both observational errors and systematic errors
in $\beta$ (see \S~\ref{sec_dist_error}).

As  stated  above,  the  massE  ruler  has  two  nuisance  parameters:
$S_{D_{0}}$ calibrates  the absolute distance, which  degenerates with
the  local  Hubble  constant,  and  $\beta$  determines  the  relative
distances  among  galaxies  with   different  masses.  The  latter  is
transformed to the curvature of the distance-redshift diagram, because
for a  flux limited  sample galaxies at  higher redshifts  have higher
masses.   The massE  ruler  would  be a  viable  new  probe of  cosmic
expansion history, only if for a fixed $\beta$ the massE ruler gives a
smooth function of the distance  vs. redshift with small fluctuations,
otherwise systematic  errors unrelated  to $\beta$, e.g.,  some hidden
parameters, may be  too large for the massE ruler  to be feasible. The
latter is in fact the case for other galaxy scaling laws.  As a sanity
check, we  let $\beta$  uniformly distribute over  a very  large range
from 0 to 1, and carry  out the fitting with flat $\Lambda$CDM through
\texttt{cosmosis}      \citep{Zuntz15}      and     emcee      sampler
\citep{Foreman-Mackey13}.  As listed in Table~\ref{tab_cosmic_result},
$\Omega_{\Lambda}$  converges   to  be   a  reasonable   value,  i.e.,
\OLambdamassEbetafree, along  with $\beta$=\betafree.   This indicates
that  once  $\beta$ is  estimated  from  the local  relationship,  the
performance of the massE ruler can further improve.

Figure~\ref{lnDA_vs_z_SDSS}  (a) shows  the derived  distance-redshift
diagram  in  terms  of  ln($D_{\rm A}/S_{D_{0}}$)  as  a  function  of
redshift,  which is  also  listed  in Table~\ref{tab_dist_result}.  As
shown   in   Figure~\ref{covariance_matrix}    (a)   and   listed   in
Table~\ref{tab_cov_matrix},  the   covariance  matrix   of  ln($D_{\rm
  A}/S_{D_{0}}$) includes two  parts.  The first one is  caused by the
observational errors  of $\sigma_{\rm e}$, $R_{\rm  e,as}$ and $M_{\rm
  star}$.  In the case of no observational errors, the derived $D_{\rm
  A}$ distribution should be  the redshift distribution convolved with
peculiar      velocity.      As      an      example     shown      in
Figure~\ref{covariance_matrix} (b)  for $z$=[0.17, 0.18],  the derived
ln($D_{\rm A}$) distribution (blue  filled) scatters much more broadly
than that implied  by the redshift distribution  (orange filled).  The
distribution can be  approximated with a central Gaussian  core plus a
broad wing. The standard deviation of  the central core is 0.29, which
is consistent  with but slightly  larger than  the square root  of the
quadratic  sum of  quoted median  errors of  three observables  in the
catalog  (red  open  symbols)  through  Equation~\ref{distance_ruler},
which gives 0.21. This indicates that the quoted errors of observables
are slightly  under-estimated.  The broad  wing is due to  outliers of
measurements.  As  a result,  we adopt the  standard deviation  of the
whole distribution divided by the square root of the number of objects
as the error. It is about two times larger than the one derived by the
bootstrap  method.   Since  each  redshift  bin  is  independent,  all
non-diagonal elements  of the  covariance matrix are  set to  zero for
this first part.  The second part  is due to the variation in $\beta$.
We adopt its measurement and associated error from the local sample to
calculate   the   corresponding   covariance  matrix   of   ln($D_{\rm
  A}/S_{D_{0}}$).  Note  that all uncertainties, e.g.,  the systematic
uncertainties in the  stellar initial mass function,  which shifts all
distances by the same amount, are taken into account by $S_{D_{0}}$.

\begin{table*}
\caption{\label{tab_cosmic_result} The best-fit cosmological parameters.}
\begin{tabular}{llllllllllllll}
\hline
   $\beta$ priors$^{\dagger}$    & $M_{\rm star}$ for massE       & rulers           &      $\beta$            & $\Omega_{\rm DE}$                 & $w$  \\
\hline

\multicolumn{6}{c}{flat $\Lambda$CDM}\\
\hline
U(0, 1)                        &  MPA-JHU                  & massE                  &  \betafree             &  \OLambdamassEbetafree                   &  -- \\
fixed to N (\betaLocalRel)     &  MPA-JHU                  & massE                  &   --                   &  \OLambdamassEbetacovary                 &  -- \\ 
fixed to N (\betaLocalRel)     &  \texttt{CIGALE}          & massE                  &   --                   &  \OLambdamassEbetacovaryPCIGALE          & -- \\  
\hline
\multicolumn{6}{c}{flat $w$CDM}\\
\hline
                      &    &    SN Ia      &   --   & 0.68$^{+0.07}_{-0.08}$  &   -1.06$^{+0.23}_{-0.19}$  \\ [4.5pt]
                      &    &    BAO        &   --   & 0.72$^{+0.02}_{-0.03}$  &   -0.73$^{+0.17}_{-0.16}$  \\ [4.5pt]
                      &    &    CMB        &   --   & 0.80$^{+0.06}_{-0.03}$  &   -1.49$^{+0.26}_{-0.31}$  \\ [4.5pt]
                      &    &    massE      &   --   & 0.67$^{+0.27}_{-0.17}$  &   -0.97$^{+0.42}_{-0.30}$  \\ [4.5pt]
\hline
\hline
\end{tabular}\\
$^{\dagger}$U stands for a uniform distribution and N is for a normal distribution. All other priors are included in Table~\ref{tab_priors}.
\end{table*}

\section{Constraints on  cosmological  parameters}

\subsection{Constraints on the flat $\Lambda$CDM}\label{sec_flatCDM}

We carry  out the fitting to  the distance vs.  redshift  diagram in
Figure~\ref{lnDA_vs_z_SDSS}  as  obtained by the massE  ruler  with a  flat
$\Lambda$CDM.    The  fitting   was  done   through  \texttt{cosmosis}
\citep{Zuntz15} with the emcee sampler \citep{Foreman-Mackey13},
whose  priors  are  listed  in Table~\ref{tab_priors} where both $h_{0}$ and $S_{D_{0}}$ are set to be
free.   As  shown  in
Figure~\ref{lnDA_vs_z_SDSS}   (c),   the   convergent   results   have
$\Omega_{\Lambda}$=\OLambdamassEbetacovary,    which     is    1-$\sigma$
consistent  with  constraints by  BAO  \citep{Ross15,  Alam17}, SN  Ia
\citep{Scolnic18} and CMB \citep{Planck20}.   The massE ruler thus proves
the  acceleration  of  the   universe  at  8.5-$\sigma$\update\,  with
elliptical  galaxies  in  the  SDSS  main  galaxy  sample  under  flat
$\Lambda$CDM.

Among  three observables,  $M_{\rm star,1Mpc}$  relies on  the stellar
population synthetic  model, while  the other two  are straightforward
measurements  with   weak  model  dependence.  We   thus  carry  out
additional  sets of  stellar  mass measurements  and  compared to  the
result of the MPA-JHU catalog.  It is found that with multi-band broad
SED,  the systematic  effect  of $M_{\rm  star}$  measurements on  the
derived  cosmological parameters  is small  as long  as the  parameter
space of  the population  synthetic model  is reasonably  sampled.  We
use \texttt{CIGALE} \citep{Boquien19} to  construct a two exponential
decaying star formation history with the second one for the late-epoch
burst. The  fraction of models  that experiences late bursts  is 20\%,
smaller than 50\% in  MPA-JHU \citep{Kauffmann03}.  The metallicity is
uniformly distributed among  four discrete  values that  are 0.2,  0.4, 1.0  and 2.5
solar, while  in MPA-JHU the  metallicity is interpolated to  a higher
resolution.  The interstellar medium extinction is uniform between 0.0
and 1.5, while the MPA-JHU catalog extends to higher values.  Although
the standard deviation of the  difference between MPA-JHU' and CICALE'
stellar masses  reaches 20\%, the  systematic offset is  much smaller.
As   listed   in    Table~\ref{tab_cosmic_result},   the   best-fit
$\Omega_{\rm \Lambda}$  is \OLambdamassEbetacovaryPCIGALE\, for  the flat
$\Lambda$CDM,   which   is   essentially   the   same   as   the   one
(\OLambdamassEbetacovary) with the MPA-JHU mass.

\subsection{The statistical power in distance  measure}\label{sec_dist_error}

To illustrate the  statistical advantage in distance  measure with the
massE ruler, Figure~\ref{comp_distance_error}  compares the fractional
error of the massE-based  distance to those by SN Ia  and BAO.  We sum
the covariance matrix to obtain this  error.  In our redshift range of
[0.05, 0.2] with a median redshift $z_{\rm med}$=0.11, the corresponding angular diameter distance
is ln($D_{\rm A}(z_{\rm med})/S_{D_{0}}$[Mpc])=\medianlnDA. This fractional distance error of
\lnDErrcomb\, is a quadratic  sum of \lnDErrStat\, by  errors of
three observables and \lnDErrBeta\, by $\beta$ uncertainties. A double
check of this  estimate can be obtained by looking  at the residual of
the best  fit as shown  in Figure~\ref{lnDA_vs_z_SDSS} (b).   For each
bin, the deviation from the true distance is caused by the statistical
error of three observables plus one realization of the intrinsic massE
relationship that  is independent from  other bins.  As a  result, the
fluctuation in the residual should converge to the result by the
covariance matrix if the number of  redshift bins is large enough.  An
error of 0.35\% is derived from the standard deviation of the residual
for the median distance of the above redshift range, which is close to
the one from the covariance matrix. We further check whether the Malmquist bias can affect this error estimate
by calculating the bias for each redshift bin and treating it as additional independent error. It is found that the
final fractional distance error increases negligibly from 0.35\% to 0.36\%.

As shown  in the figure,  the measurement  with massE gives  about one
order   of  magnitude   smaller   fractional   errors  than   existing
measurements at  similar redshift  with SN  Ia and  BAO \citep{Ross15,
  Alam17,  Scolnic18}.   As  compared to  next-generation  surveys  as
offered by  BAO from  DESI and SN  Ia from  Rubin/Roman \citep{Feng14,
  DESI16}, the massE  ruler still offers 2-4  times smaller fractional
error at $z$  $\sim$ 0.1.  This demonstrates the  statistical power of
the massE ruler as a new probe of cosmic geometry.

The  statistical power  of the  massE ruler  can be  understood in  the
following  way.  Although  each SN Ia  offers a fractional distance  error of $\sim$ 5\% as
compared to $\sim$ 25\% of the massE, the number of elliptical galaxies
is far more numerous than the SN Ia. On the other hand, BAO  is  essentially
volume-limited  at low-$z$.

\subsection{Constraints on the dark energy equation  of state CDM ($w$CDM)}

The  low-$z$  distance   measurements  at  high  accuracy   by  massE  is
particularly useful  to probe the  deviation from the constant  of the
dark energy density that dominates  at the low-$z$ universe.  As shown
in Figure~\ref{w_omegam_DiffProbers},  we illustrate this  by carrying
out constraints  on the $w$CDM model  with the massE data  and compare to
those   with  the   Pantheon   SN  Ia   data  \citep{Scolnic18},   the
MGS/BOSS/eBOSS BAO data \citep{Ross15,  Alam17, Alam21} and the Planck
2018  CMB  TT,TE,EE+lowE  data  \citep{Planck20},  respectively.   All
fittings have been done with \texttt{cosmosis}   \citep{Zuntz15} and emcee
sampler \citep{Foreman-Mackey13}.       As       listed      in 
Table~\ref{tab_cosmic_result}, the  massE alone constrains  $w$ to  an error
that   is   only  a   factor   of   \wimproveBAO,  \wimproveSN\,   and
\wimproveCMB\, larger than the current one  by BAO only, SN Ia only and
CMB only, respectively.




\section{Conclusion}

In this  study we have shown  that the massE relationship  of galaxies
could be  a new  cosmic ruler to  probe the  distance-redshift diagram
with  advantage  of  its  statistical  power.   It  has  two  nuisance
parameters  with three  observables  for ellipticals  that are  galaxy
sizes,  velocity dispersions  and stellar  masses. With  the SDSS  MGS
sample, the distance  at $z$=0.11 with the massE  ruler is constrained
to a fractional error of \lnDErrcomb, with the  best-fit dark energy
   density of \OLambdamassEbetacovary\, for flat $\Lambda$CDM. In principle the massE can be
applied to higher $z$ by combining large-area space-based imaging survey
and ground-based deep spectroscopic survey. 

The
\texttt{cosmosis} modules including the input parameter files and data
files     for    the     massE   can be obtained through the link available
on  arXiv of this manuscript.

\section*{Acknowledgements}

We thank the  referee for constructive comments that  help improve the
paper. We thank Pengjie Zhang for helpful comments. Y.S.  acknowledges
the support from the National  Key Research and Development Program of
China (No.   2018YFA0404502), the National Natural  Science Foundation
of  China (NSFC  grants 11825302,  12141301, 12121003,  11733002), the
science  research grants  from  the China  Manned  Space Project  with
NO. CMS-CSST-2021-B02, and the  Tencent Foundation through the XPLORER
PRIZE.   S.M. is  partly supported  by the  National Key  Research and
Development Program  of China  (No.  2018YFA0404501), by  the National
Natural  Science  Foundation  of   China  (11821303,  11761131004  and
11761141012) and by Tsinghua University Initiative Scientific Research
Program  ID  2019Z07L02017.  Z.Y.Z  acknowledges the  support  of  the
National Natural Science  Foundation of China (NSFC)  under grants No.
12041305, 12173016,  the Program for Innovative  Talents, Entrepreneur
in  Jiangsu, and  the science  research grants  from the  China Manned
Space Project with NO.  CMS-CSST-2021-A08, CMS-CSST-2021-A07.

Funding for  the SDSS and SDSS-II  has been provided by  the Alfred P.
Sloan Foundation, the Participating Institutions, the National Science
Foundation, the  U.S.  Department of Energy,  the National Aeronautics
and Space Administration, the  Japanese Monbukagakusho, the Max Planck
Society, and  the Higher  Education Funding  Council for  England. The
SDSS  Web Site  is http://www.sdss.org/.  The SDSS  is managed  by the
Astrophysical    Research    Consortium    for    the    Participating
Institutions. The  Participating Institutions are the  American Museum
of  Natural History,  Astrophysical Institute  Potsdam, University  of
Basel,  University  of  Cambridge, Case  Western  Reserve  University,
University of Chicago, Drexel  University, Fermilab, the Institute for
Advanced  Study,   the  Japan   Participation  Group,   Johns  Hopkins
University, the  Joint Institute  for Nuclear Astrophysics,  the Kavli
Institute  for   Particle  Astrophysics  and  Cosmology,   the  Korean
Scientist Group, the Chinese Academy  of Sciences (LAMOST), Los Alamos
National  Laboratory, the  Max-Planck-Institute for  Astronomy (MPIA),
the  Max-Planck-Institute for  Astrophysics  (MPA),  New Mexico  State
University,   Ohio  State   University,   University  of   Pittsburgh,
University  of Portsmouth,  Princeton  University,  the United  States
Naval Observatory, and the University of Washington.

\section*{Data Availability}

All the  data used  here are available  upon reasonable  request.



\bibliographystyle{mnras}
\bibliography{ms} 




\appendix

\section{Priors of cosmological parameters}\label{appendix_prior}
The priors of cosmological parameters are detailed in Table~\ref{tab_priors}.
  All fittings have been done with \texttt{cosmosis}   \citep{Zuntz15} and emcee
sampler \citep{Foreman-Mackey13}. Each fitting has 4$\times$10$^{6}$ samples, with 
the Gelman-Rubin (G-R) convergence value of $<$ 0.01, except for CMB that has
15$\times$10$^{6}$ samples in order to have G-R value below 0.01.

\begin{table}
\caption{\label{tab_priors} Priors of cosmological parameters.}
\begin{tabular}{llllllllllllll}
\hline
parameters       & priors & models   \\
\hline
\hline
\multicolumn{3}{c}{massE only} \\
\hline
$\Omega_{\rm m}$  &   U(0.0,  1.0)     & flat $\Lambda$CDM, flat $w$CMD \\
$h_{0}$          &  U(0.2, 1.0)      & flat $\Lambda$CDM, flat $w$CMD    \\
$S_{D_{o}}$       &  U(0.3,  2.0)     & flat $\Lambda$CDM, flat $w$CMD \\
$w$              &  U(-3.0, 0.0)    &  flat $w$CMD \\
\hline
\hline
\multicolumn{3}{c}{SN Ia only} \\
\hline
$\Omega_{\rm m}$   &   U(0.0,  1.0)     & flat $w$CMD \\
$h_{0}$           &  U(0.2, 1.0)      &  flat $w$CMD    \\               
$M$              & U(-23, -15)       &  flat $w$CMD    \\       
$w$              &  U(-3.0, 0.0)    &  flat $w$CMD \\
\hline
\hline
\multicolumn{3}{c}{BAO only} \\
\hline
$\Omega_{\rm m}$   &   U(0.0,  1.0)       & flat $w$CMD \\
$h_{0}$            &  U(0.2,   1.0)      &  flat $w$CMD    \\               
$\Omega_{\rm b}$   &   U(0.001, 0.3)       &  flat $w$CMD    \\       
$w$               &  U(-3.0, 0.0)        &  flat $w$CMD \\
\hline
\hline
\multicolumn{3}{c}{CMB only} \\
\hline
$\Omega_{\rm c}$   &   U(0.0, 0.9)          & flat $w$CMD  \\
$\Omega_{\rm b}$   &   U(0.0, 0.3)         &  flat $w$CMD    \\       
$h_{0}$           &   U(0.2, 1.0)          & flat $w$CMD  \\
$w$               &  U(-3.0, 0.0)          &  flat $w$CMD \\
$n_{s}$           & U(0.8, 1.2)             & flat $w$CMD     \\
$10^{9}A_{s}$       & U(1.48, 5.45)           &  flat $w$CMD    \\
$k_{s}$            & 0.05                  & flat $w$CMD     \\
$\tau$             & U(0.01, 0.8)           &  flat $w$CMD    \\
\hline
\hline
\end{tabular}\\
U stands for a uniform distribution.

\end{table}



\bsp	
\label{lastpage}
\end{document}